\documentclass{article}
\pdfoutput=1

\usepackage[final,nonatbib]{neurips_2022_ml4ps}

\usepackage[utf8]{inputenc} 
\usepackage[T1]{fontenc}    
\usepackage{hyperref}       
\usepackage{url}            
\usepackage{booktabs}       
\usepackage{amsfonts}       
\usepackage{nicefrac}       
\usepackage{microtype}      
\usepackage{xcolor}         
\usepackage{amsmath}
\usepackage{graphicx}
\usepackage{wrapfig}

\newcommand{\be}{\begin{equation}}
\newcommand{\ee}{\end{equation}}

\title{A fast and flexible machine learning approach\\ to data quality monitoring }

\author{%
  Gaia Grosso \\
  Dipartimento di Fisica e Astronomia \\
  Università di Padova \\
  INFN,  Sez. di Padova \\
  Padova,  Italy \\
  CERN, Experimental Physics Department \\
  Geneva, Switzerland \\
  \texttt{gaia.grosso@cern.ch} \\
   \And
   Nicolò Lai\\
   Dipartimento di Fisica e Astronomia \\
   Università di Padova \\
   Padova,  Italy \\
   \texttt{nicolo.lai@studenti.unipd.it} \\
   \And
   Marco Letizia \\
   MaLGa Center - DIBRIS \\
   Università di Genova \\
   INFN, Sez. di Genova \\
   Genova, Italy \\
  \texttt{marco.letizia@edu.unige.it} \\
   \And
   Jacopo Pazzini \\
   Dipartimento di Fisica e Astronomia \\
   Università di Padova \\
   INFN,  Sez. di Padova \\
   Padova,  Italy \\
   \texttt{jacopo.pazzini@unipd.it} \\
   \And
   Marco Rando \\
   MaLGa Center - DIBRIS \\
   Università di Genova \\
   Genova, Italy \\
   \texttt{marco.rando@edu.unige.it} \\
   \And
   Andrea Wulzer \\
   Dipartimento di Fisica e Astronomia \\
   Università di Padova \\
   Padova,  Italy \\
   \texttt{andrea.wulzer@cern.ch} \\
   \And
   Marco Zanetti \\
   Dipartimento di Fisica e Astronomia \\
   Università di Padova \\
   INFN,  Sez. di Padova \\
   Padova,  Italy \\
   \texttt{Marco.Zanetti@cern.ch} \\
}

\begin{document}

\maketitle

\begin{abstract}
    We present a machine learning based approach for real-time monitoring of particle detectors. The proposed strategy evaluates the compatibility between incoming batches of experimental data and a reference sample representing the data behavior in normal conditions by implementing a likelihood-ratio hypothesis test. The core model is powered by recent large-scale implementations of kernel methods, nonparametric learning algorithms that can approximate any continuous function given enough data. The resulting algorithm is fast, efficient and agnostic about the type of potential anomaly in the data. We show the performance of the model on multivariate data from a drift tube chambers muon detector.
\end{abstract}

\section{Introduction}
\label{sec:intro}
Modern high-energy physics experiments consist of complex detectors where hundreds of millions of sensors are read out as frequently as every few nanoseconds.
The electrical signals are amplified, processed and combined before the trigger selection and the final storage. At each of these steps some errors can occur and invalidate the whole process. Monitoring systems are deployed to assess the quality of the data and keep the flow under control throughout all the stages of the acquisition.
Data quality monitoring (DQM) is a challenging task from a statistical point of view due to its high intrinsic dimensionality and the high level of human supervision required.
The unforeseen events incoming during an experimental run can be several. Some of them can be anticipated and recognised while they occur, others cannot. The detection of well known dysfunctions could be nonetheless missed due to the huge amount of channels that should be simultaneously monitored. Developing flexible highly automatized techniques to supervise multiple variables at once is thus fundamental to reduce the risk of failures \cite{pol2022data,pol2019detector,Azzolini:2701776,adinolfi2017lhcb}.
This work proposes the use of a recent machine learning approach to compare collected data with a sample of reference events that depicts the correct detector readings. This reference sample can be, for instance, a set of measurements in a controlled scenario or simulated events. The basic idea is to perform a hypothesis test powered by a fast and flexible machine learning (ML) algorithm. In practice, we leverage the ability of binary classifiers to implicitly model the underlying data-generating distributions and estimate the likelihood ratio test statistics. This is then used to assess whether the hypothesis underlying the observed data (alternative hypothesis) agrees with the assumption of normal behavior (null hypothesis). 
If a set of measurements significantly deviates from the reference sample,  the learned likelihood ratio can be used to characterize the anomalies in the feature space. To increase the model's sensitivity, we take a large reference sample and reduce its statistical fluctuations. This work is inspired by the ML model introduced in Ref.~\cite{Letizia:2022xbe} in the context of model-independent new physics searches in high-energy physics. In that work, the authors propose an algorithm based on kernel methods with clear advantages compared to similar neural network implementations in terms of training time while obtaining similar performance. This feature makes the kernel approach ideal for DQM,  where fast training is essential for online analyses. \\ 
The paper is organized as follows. In the next section we introduce the experimental setup and the algorithm input variables. These include a reference data set collected under standard conditions and smaller samples with anomalous controlled behaviors. The ML model and our core strategy are then described in Section \ref{sec:model}, whereas an overview of the results is given in Section \ref{sec:results}. Finally, the last section is devoted to conclusions and further developments.

\section{Experimental setup and data samples}
\label{sec:setup}
\begin{wrapfigure}{r}{0.58\textwidth}
\centering
\includegraphics[width=0.30\textwidth]{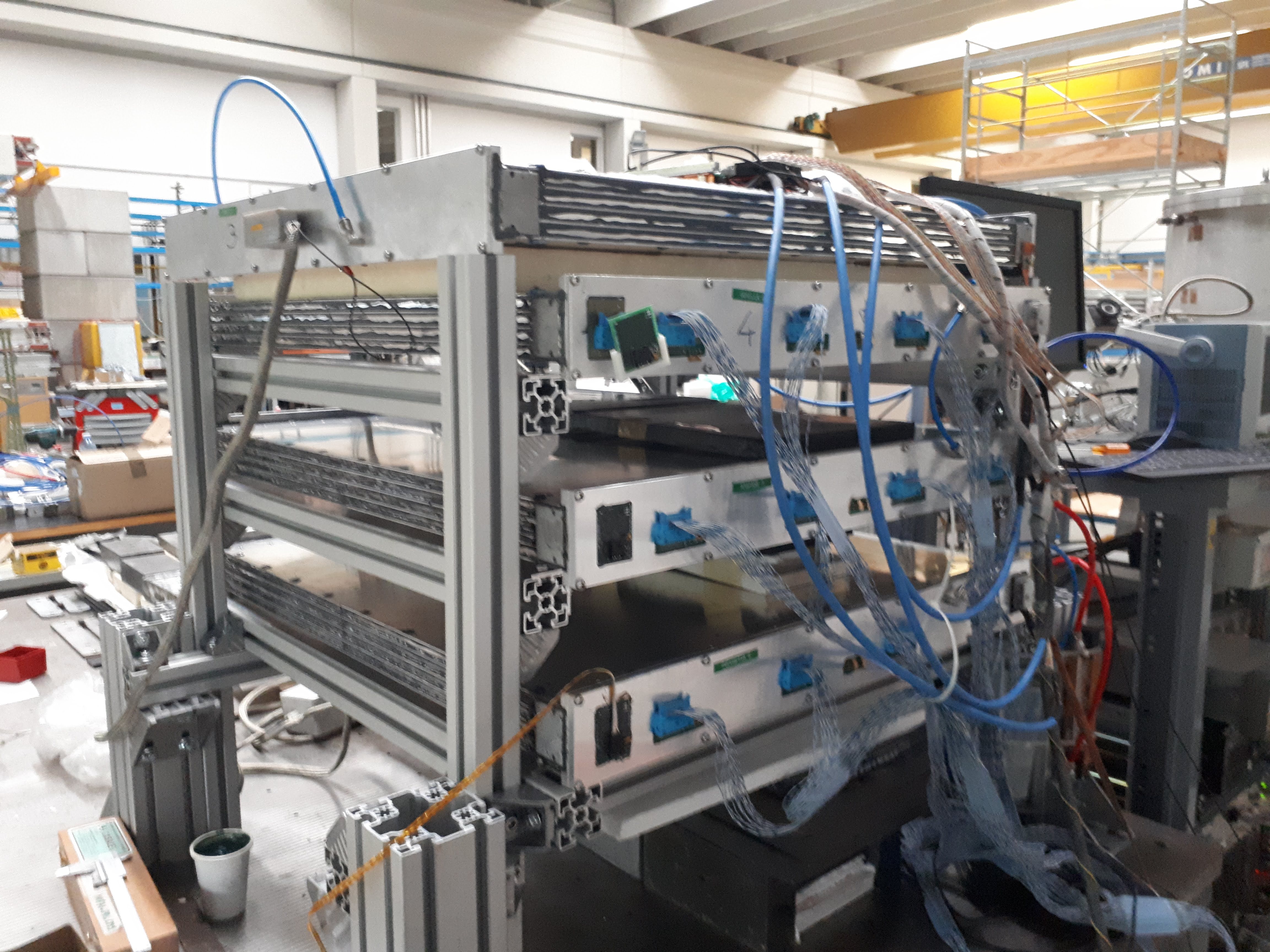}
\includegraphics[width=0.27\textwidth]{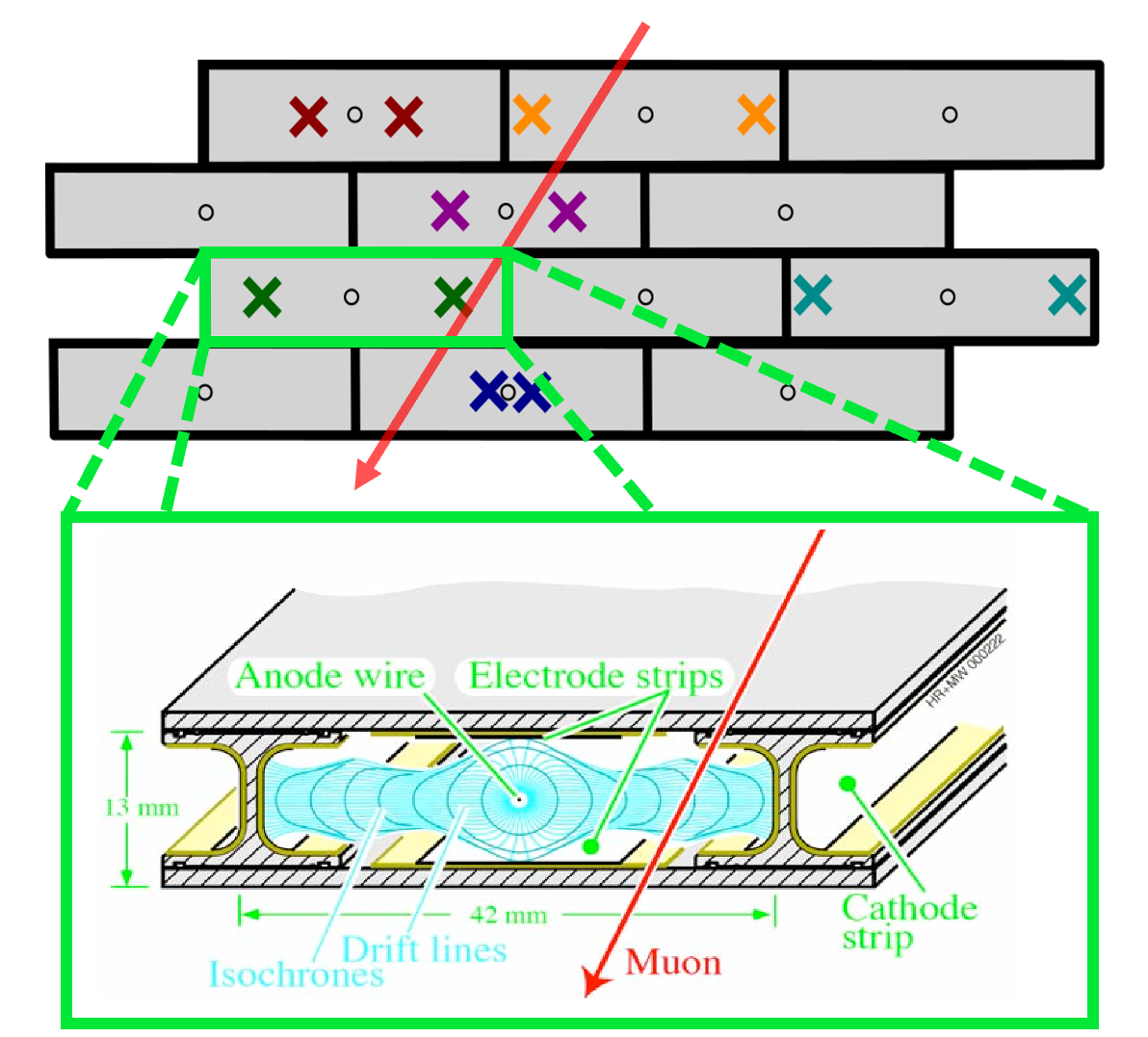}
\caption{Left: the experimental apparatus at Legnaro Laboratory, with four drift-tube chambers vertically stacked. 
Right: a schematic view of the cell (bottom) and the hit pattern left by a charged particle crossing (top).}
\label{apparatus}
\end{wrapfigure}

In this work we consider an experimental apparatus consisting of a set of drift tube (DT) chambers developed and installed at the Legnaro INFN National Laboratory (Fig.\ref{apparatus}, left). It represents a reduced-scale version of the ones installed in the CMS experiment at the LHC~\cite{CMS:2008xjf}.
The fundamental element of the detector is a $70$ cm long tube with a $4\times2.1$ ${\rm cm^2}$ cross section (Fig.\ref{apparatus}, bottom right). Within each tube, an electric field is produced by an anodic wire ($+3.6$ kV) laid in the center and two cathodic strips ($-1.2$ kV) at the sides;  an additional pair of strips at an intermediate voltage ($+1.8$ kV) are placed above and below the wire to improve the field homogeneity.
The tubes are filled with a ${\rm Ar-CO_2}$ gas mixture ($85\%-15\%$) whose molecules are ionized when charged particles traverse the sensitive volume. 
The electrons produced ionizing the gas drift with constant velocity along the field lines towards the wire, where they are collected. The arrival time is recorded by the front-end electronics, which amplify the signal and filter noise below a specific threshold (nominal at 100 mV). 
The particle's position (with a left-right ambiguity) is linearly dependent on the drift time. The two parameters of the linear relation are the drift velocity, known after calibration, and the time pedestal, which can be provided by an external trigger or by means of a \textit{mean-timer} technique \cite{Migliorini:2021fuj}.\\
A drift tube chamber comprises 64 tubes grouped into four layers of 16 each. The layers are staggered horizontally by half a cell. From the measurements collected from of a single chamber, it is thus possible to track the position and slope of the charged particles crossing the detector by performing a linear fit of the hits from the four layers (Fig.\ref{apparatus}, top right). The setup in Legnaro records muons from cosmic rays (at a sea-level rate of about 1 per minute per ${\rm cm^2}$) and exploits plastic scintillators as an external trigger.\footnote{More precisely, the data are acquired continuously at 40 MHz, with the trigger information from the scintillators added to the data stream}
Each detector channel provides an (absolute) time stamp once a signal is recorded. Since we are interested in muons crossing the detector, we group the hits corresponding to such events and we reconstruct the muon track out of the observed hits pattern.\\
The performances of the muons track reconstruction deteriorate as soon as the detector conditions are anomalous. 
The following list of relevant quantities can be exploited to assess the quality of the recorded data:\\
$\bullet$ Drift times ($t_i$): the four drift times associated with the muon track. \\
$\bullet$ Slope ($\phi$): the angle formed by the muon track in the plane orthogonal to the anodic wires (measurement plane) with respect to the vertical axis. \\
$\bullet$ Number of hits ($n_{hits}$): the number of hits recorded in a time window of one second around the muon crossing time. \\
To test the performances of the DQM approach presented in this work, we reproduce two of the typical failures that occur during detector operations. In particular, we consider the eventuality of the cathodic strips voltage reduced to $25\%$ of their nominal value (i.e., to $-300$ V) and that of the front-end thresholds reduced to $75$, $50$ and $25$ mV, respectively 75\%, 50\% and 25\% of the nominal value. 
A dedicated data collection campaign has been launched.\footnote{\url{https://doi.org/10.5281/zenodo.7128223}} As a result, a set of data samples has been collected, one consisting of $3\times 10^5$ events being the reference and the other four of approximately $10^4$ events, each corresponding to one of the anomalous conditions mentioned above.

\section{Methodology}
\label{sec:model}
\textbf{Model design.} To estimate the likelihood ratio,  we train a binary classifier on a reference sample $S_0=\{x_i\}_{i=1}^{n_0}$ and on a set of measurements $S_1=\{x_i\}_{i=1}^{n_1}$ using a weighted logistic loss
\be\label{wbce}
\ell(y,f(x))=\frac{n}{n_0} (1-y) \log \left(1+e^{f_w(x)}\right)+\frac{n}{n_1} y \log \left(1+e^{-f_w(x)}\right).
\ee
with $n=n_0+n_1$ and where $y$ is the class label that takes the zero value for $S_0$ and one for $S_1$.  The model parameters $w$ are going to be learned from the data.  The weights $n/n_y$ are introduced to compensate for the data imbalance $(n_0\gg n_1)$ while keeping the statistical advantage of a large reference sample.  
The learned classifier $f_{\hat{w}}$ is then obtained by minimizing the average loss over the training set.  By a standard computation, the learned function can be shown to approximate the likelihood ratio $f_{\hat{w}}\approx f^*(x)=\log\frac{p(x|1)}{p(x|0)}$ (see, for instance, appendix A of Ref.~\cite{Letizia:2022xbe}).
The log-likelihood ratio test statistics can then be easily evaluated on the measurements as
\be\label{t}
t_{\hat{w}}(S_1)=2\sum_{x\in S_1} f_{\hat{w}}(x).
\ee
In this work, we consider a learning algorithm based on kernel methods \cite{hastie01statisticallearning} of the form
$f_w(x)=\sum_{i=1}^n w_i k_\sigma(x,x_i)$, where $k_\sigma (x,x_i)=\exp{(-\Vert x-x'\Vert^2/2\sigma^2)}$ is the Gaussian kernel function and the width $\sigma$  is a hyper-parameter.  Kernel methods are universal in the large sample limit \cite{JMLR:v7:micchelli06a,christmann2008support} but they scale poorly with data size.  For this reason,  we consider here Falkon \cite{falkonlibrary2020},  a modern library which makes use of a number of algorithmic ideas,  such as column sub-sampling,  and takes full advantage of GPU hardware to extend the use of kernel methods to large scale scenarios with fast training and strong statistical guarantees.\\
\textbf{Model selection and training.} The training is performed by considering the empirical risk based on Eq.~\eqref{wbce} with an added L2 regularization term \cite{shalev2014understanding} weighted by a hyper-parameter $\lambda$. 
While the machine learning model is at its core a binary classifier,  we do not monitor typical classification metrics such as binary accuracy or the AUC,  and our metric of interest is the likelihood ratio test statistics given in Eq.~\eqref{t}. 
Hyper-parameters are tuned following the prescription given in Ref.~\cite{Letizia:2022xbe}, which includes a mix of heuristics,  statistical optimality and efficiency requirements. \\
\textbf{Anomaly detection.} The analysis strategy for the detection of anomalies in the measurements is composed of three steps:\\
$\bullet$ The distribution of the test statistics under the null (reference) hypothesis is estimated by training the model multiple times ($\mathcal{O}(100)$) on the reference sample $S_0$ and on different data samples $S_1$ also following the reference distribution, e.g., anomaly free.\\
$\bullet$ A single training is performed on the reference $S_0$ and the actual measurements $S_1$ to estimate the test statistic $t_{\hat{w}}(S_1)$ of the measurements.\\
$\bullet$ The $p$-value corresponding to $t(S_1)$ is computed with respect to the test statistic distribution under the reference hypothesis, learned in the first step.
If a statistically significant deviation from the reference data is found (for instance with respect to a threshold p-value $p^*$), the nature of the discrepancy can be analyzed in feature space by re-weighting the reference sample with the learned density ratio $\exp(f_{\hat{w}})$,  as described in the following section.\\
If a database of observations characterizing known anomalous behaviors is available, different strategies can be employed to further characterize anomalous readings. 
We tested different ML approaches for this task, among which supervised binary classifiers, multi-class classifiers and parametrized regression models. These models are trained in a fully supervised fashion to separate reference data from specific types of anomalies. This step is not discussed in this manuscript and will be presented in details elsewhere.

\section{Results}
\label{sec:results}
\begin{figure}[h!]
\centering
\includegraphics[width=0.40\textwidth]{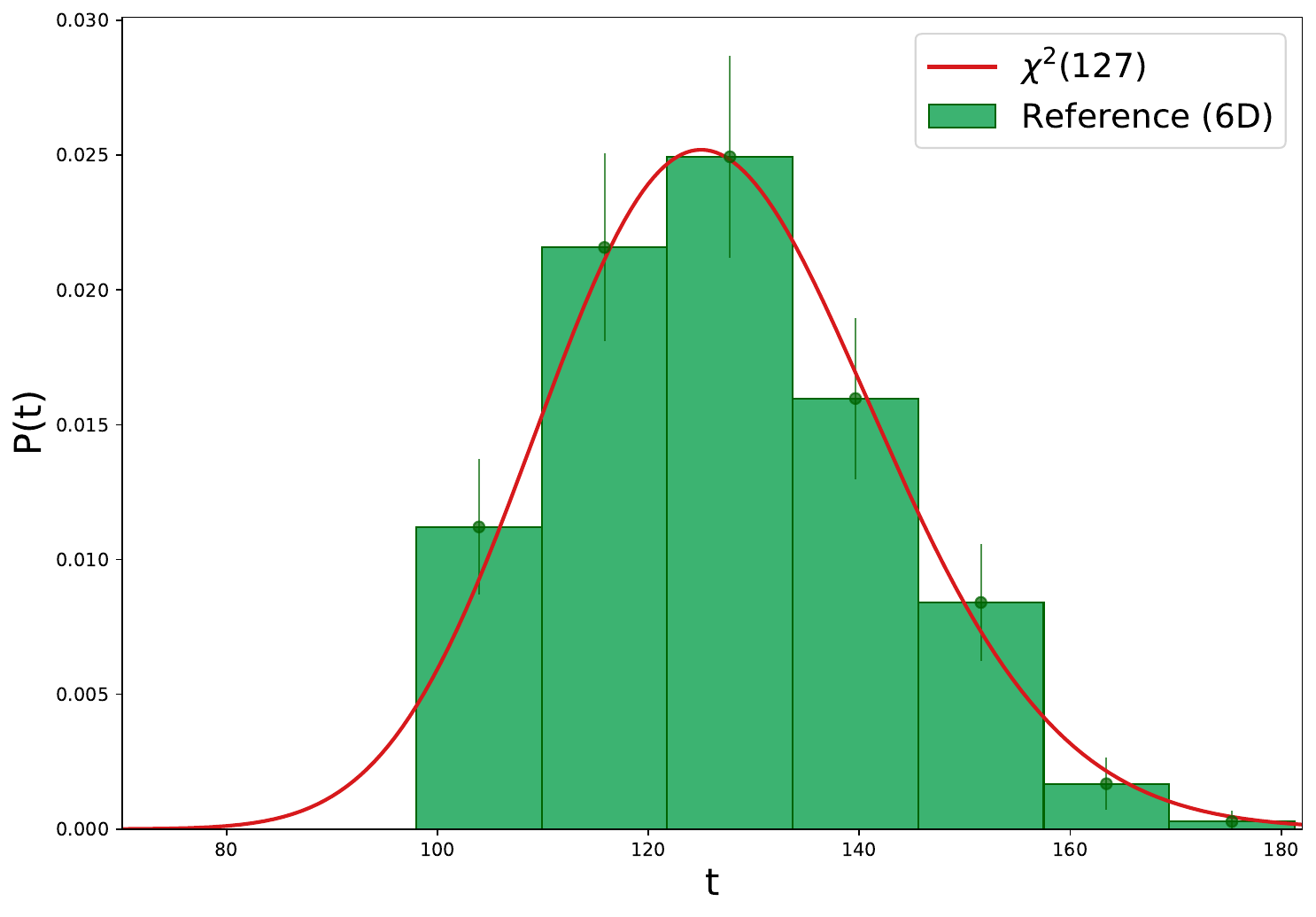}
\includegraphics[width=0.40\textwidth]{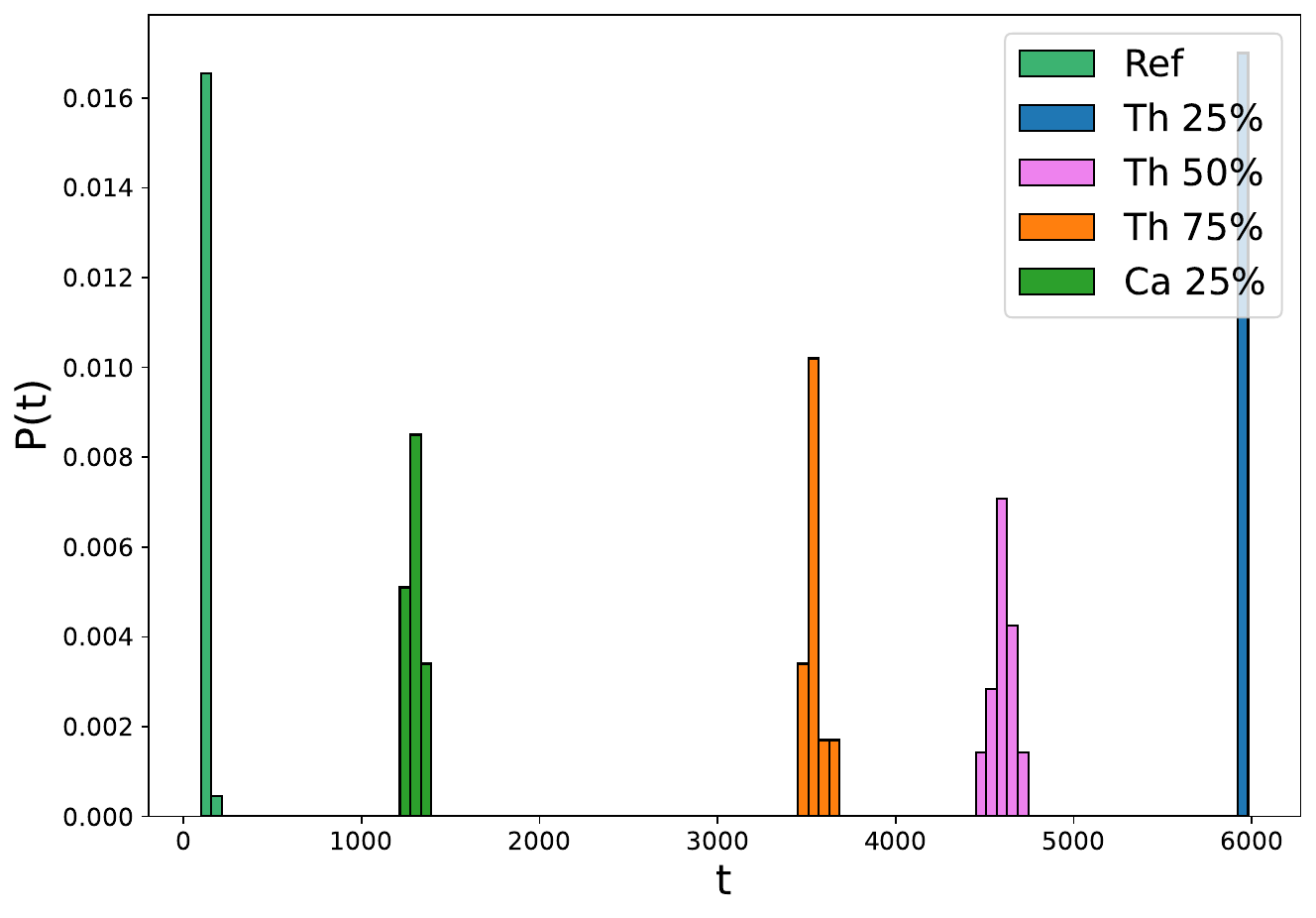}
\caption{Distribution of the test statistics under the null hypothesis and for the anomalous data in 6D.}
\label{fig:histograms6D}
\end{figure}

We present here the results obtained by applying this strategy to the monitoring of DT chambers, following the steps described in the previous section.  Having real measurements at our disposal, this study represents a simplified but realistic scenario.  As stated in \ref{sec:setup},  the input data is composed of six features (four drift times, the incoming angle, and the number of hits).
For each training, we consider a reference of $n_0=10^4$ events, sampled from the totality of the available reference data,  and a data sample of $n_1=10^3$ events (corresponding to approximately 10 minutes of data taking).  We pre-process the data by dividing each feature by the standard deviation of the reference data. The model is trained on a single machine equipped with a NVIDIA Titan Xp GPU with 12 GB of VRAM.\\
We first look at the full six-dimensional problem.  In the left-side plot of Figure \ref{fig:histograms6D},  we show the distribution of the likelihood ratio test statistics under the reference hypothesis ($S_0$ and $S_1$ both anomaly-free) and observe that it is well described by a $\chi^2$ distribution (see Ref.~\cite{Wilks:1938dza,wald1943tests,Cowan:2010js}).  In the right-side plot of Figure \ref{fig:histograms6D},  we show the values of the test statistics with different types of anomalous measurements. The model is able to separate without false negatives the different types of anomalies in a coherent way.  For instance,  as the threshold gets lower,  the data are recognized as departing more from the reference. The training time for a single classification step is approximately 1.5 seconds. We processed ten distinct data samples for each type of anomaly.
\begin{figure}[h!]
\centering
\includegraphics[width=0.40\textwidth]{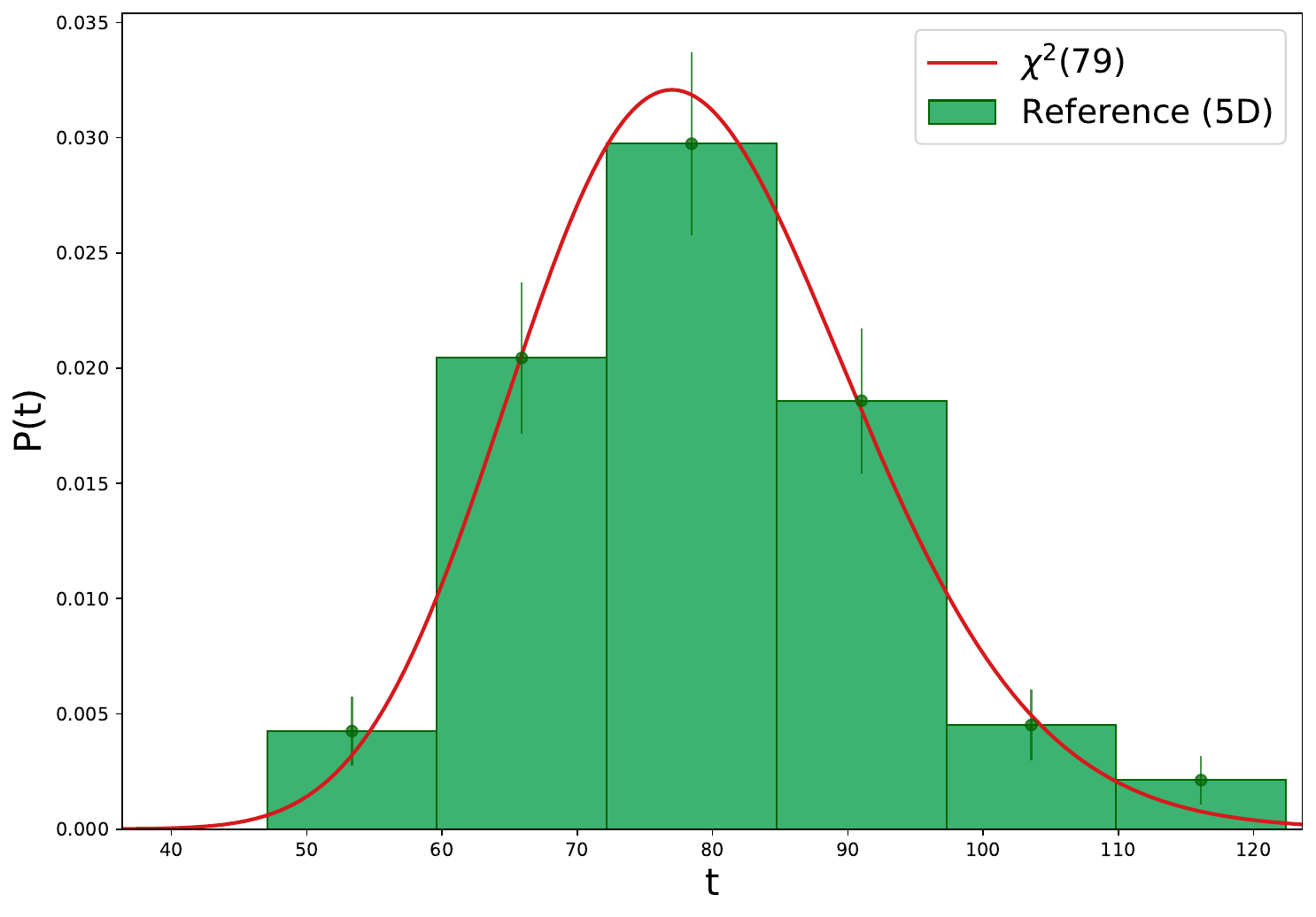}
\includegraphics[width=0.435\textwidth]{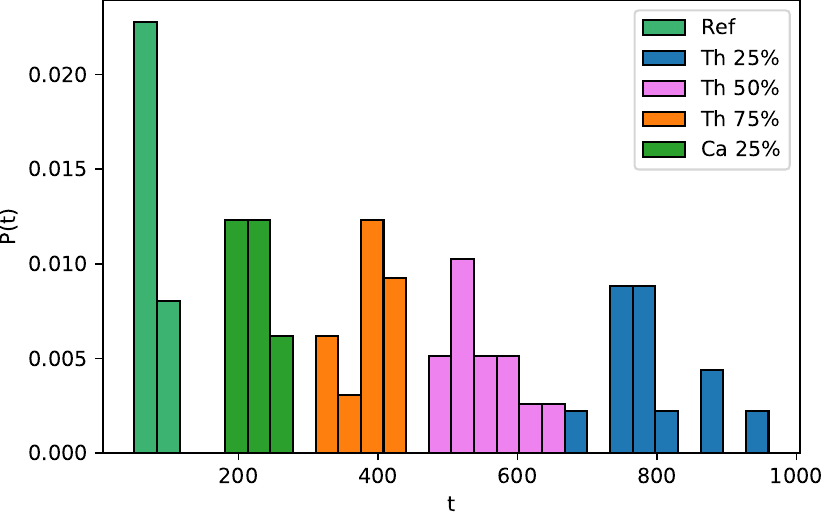}
\caption{Distribution of the test statistics under the null hypothesis and for the anomalous data in 5D.}
\label{fig:histograms5D}
\end{figure}

\begin{figure}[h]
\centering
\includegraphics[width=0.245\textwidth]{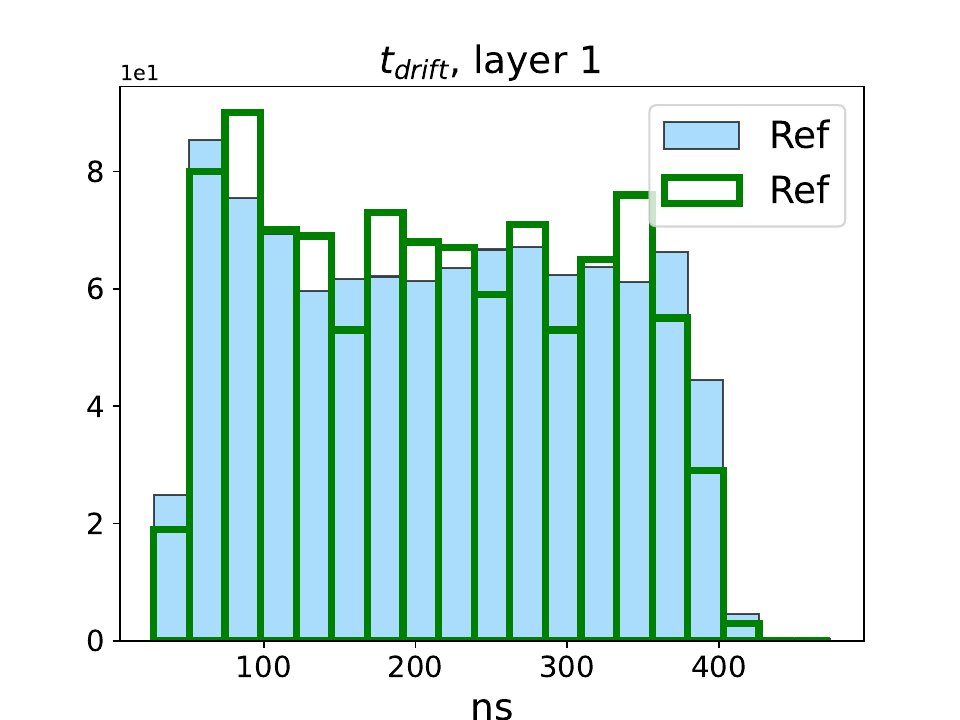}
\includegraphics[width=0.245\textwidth]{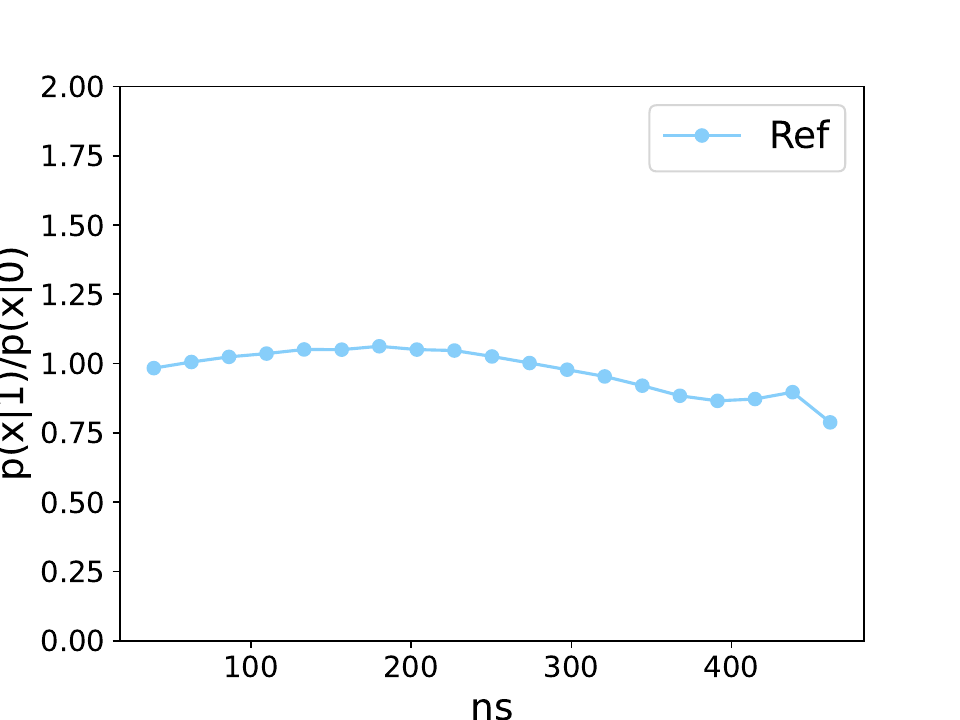}
\includegraphics[width=0.245\textwidth]{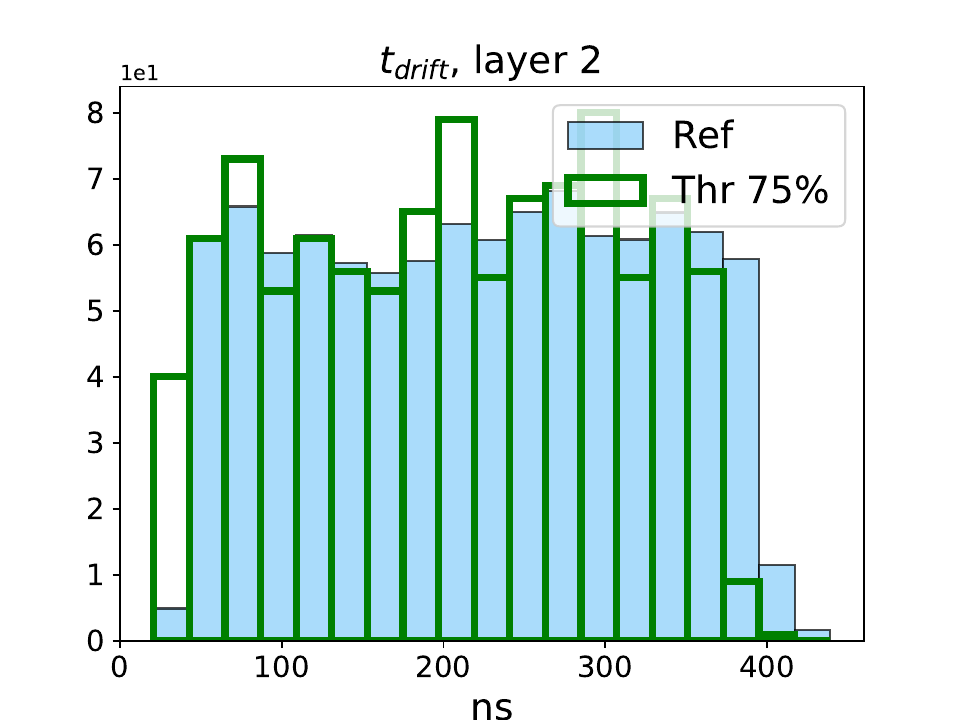}
\includegraphics[width=0.245\textwidth]{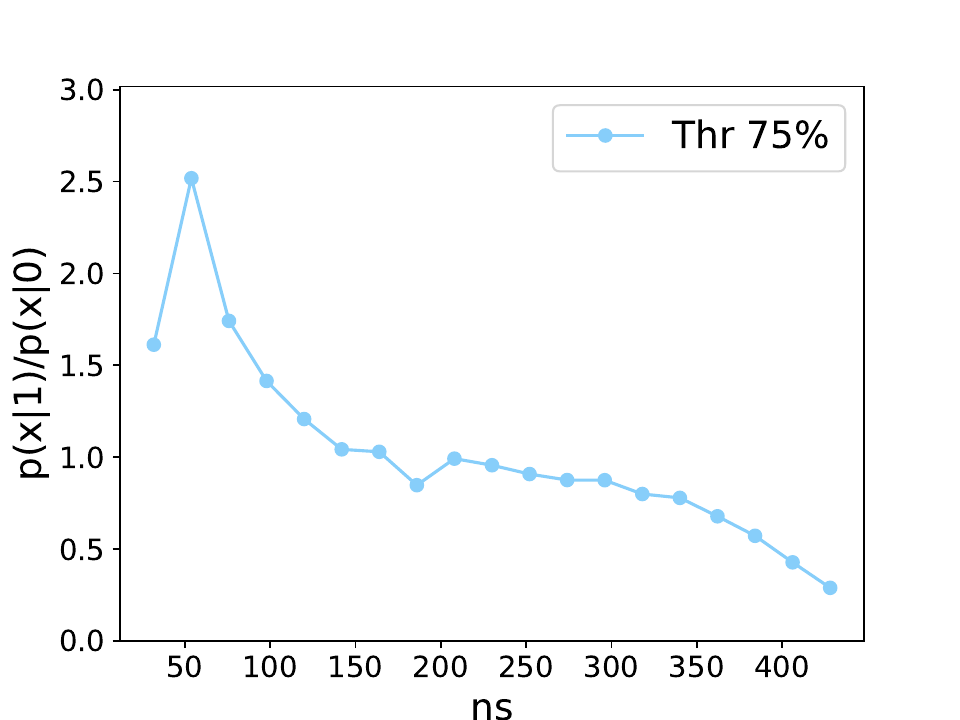}
\includegraphics[width=0.245\textwidth]{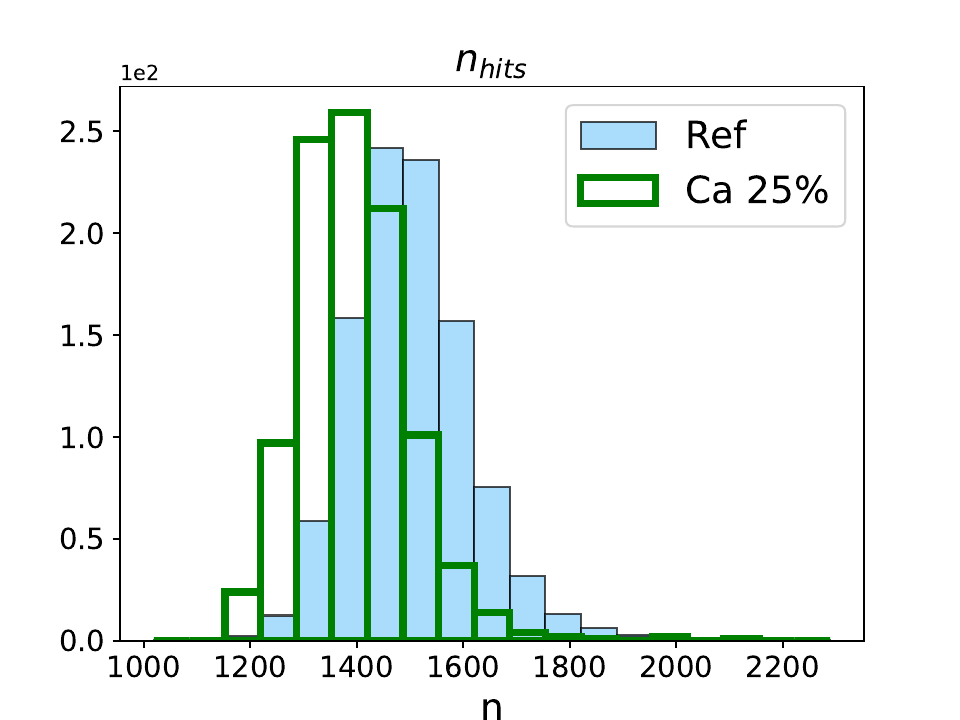}
\includegraphics[width=0.245\textwidth]{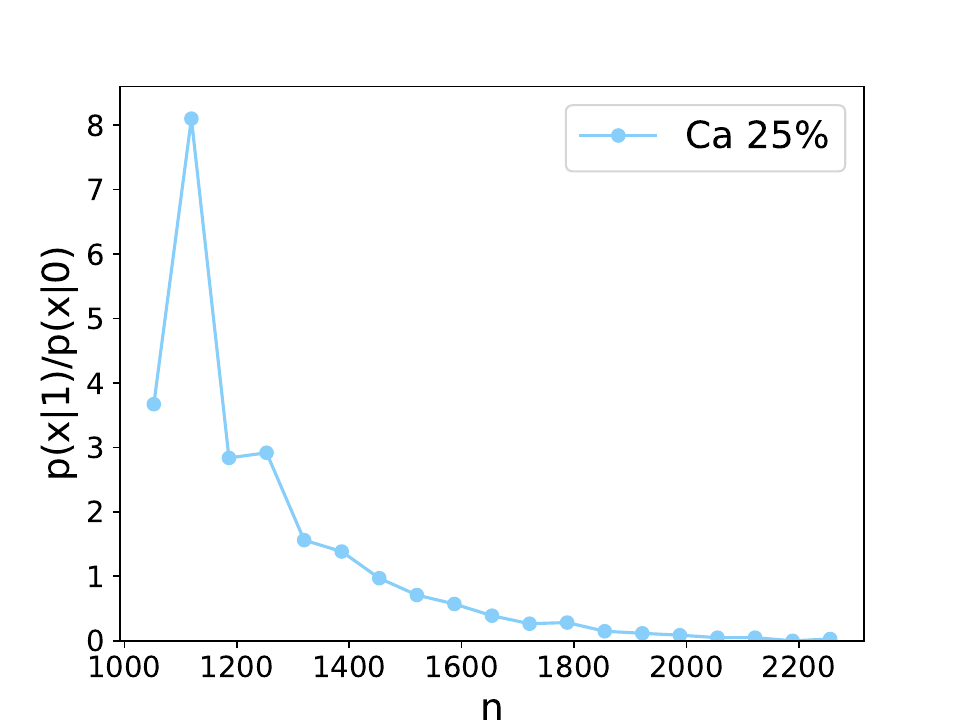}
\includegraphics[width=0.245\textwidth]{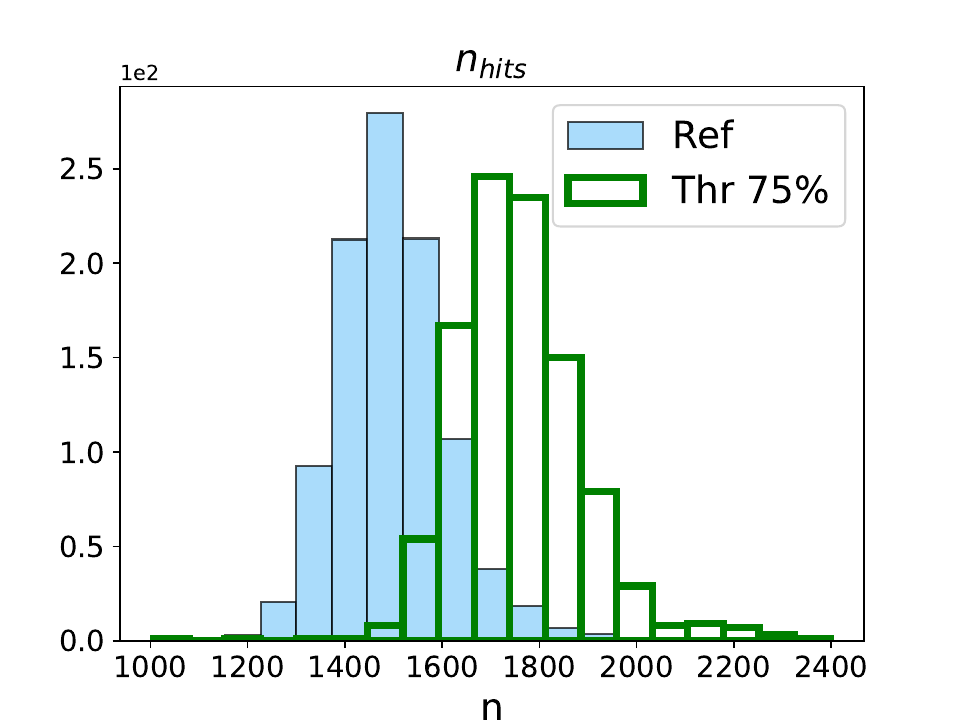}
\includegraphics[width=0.245\textwidth]{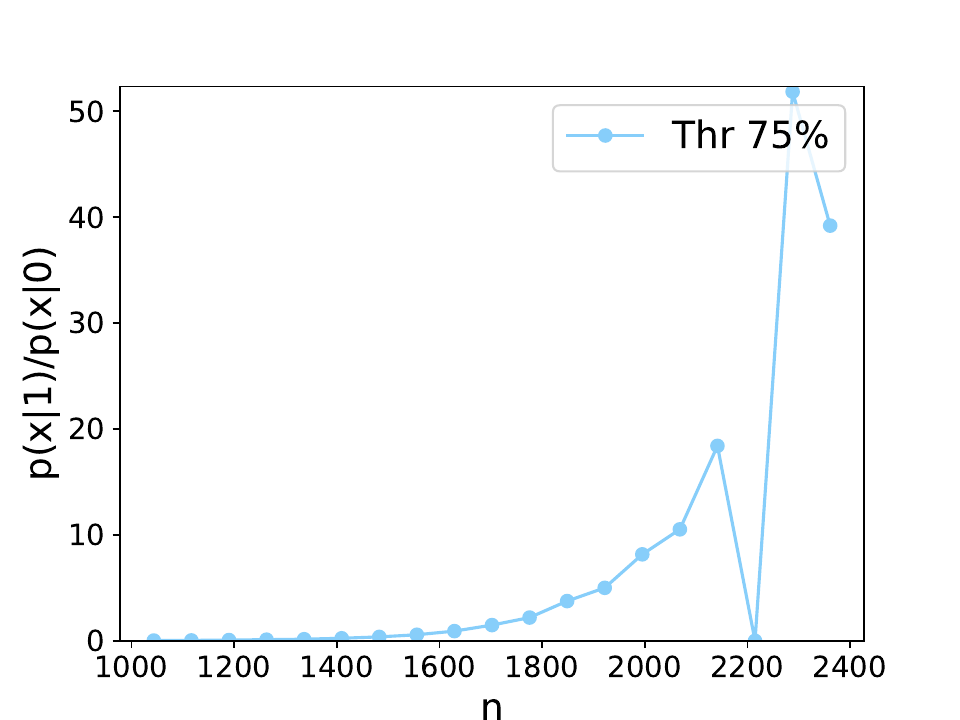}
\caption{Examples of input data and respective learned likelihood ratios. Upper plots are produced without including $n_{hits}$ among the input features.}\label{fig:outputs}
\end{figure}

We now want to test the model when a given feature, which results to be particularly discriminant with respect to certain anomalies, is not provided as input.  That is the case for $n_{hits}$ with respect to the threshold anomalies, hence we remove it and perform new experiments in five dimensions. The distribution of the test statistics is reported in Figure \ref{fig:histograms5D}. We see that the model is still able to successfully separate the various cases. In this case, the training time is reduced to approximately 0.5 seconds. To give an approximate but quantitative measure of the worst-case scenario, we estimate the p-value associated with the anomalous data sample closest to the reference distribution in 5D. By using a $\chi^2$ null distribution (left-side plot in Figure \ref{fig:histograms5D}) 
we obtain $p\approx 10^{-11}$.\\
In Figure \ref{fig:outputs},  we show instances of how to characterize potentially anomalous data by looking at the learned likelihood ratio.  These plots are produced by re-weighting the features of the reference sample by $e^{f_{\hat{w}}(x)}$ (evaluated on the reference training data),  binning it and taking the ratio with the same binned reference sample (unweighted).\footnote{We point out that the result of this operation is expected to be close but not identical to taking the ratio of the binned marginal distributions.}  This quantity is expected to be approximately one if no deviations are found in the measurements compared to the reference.  We see that the model is able to successfully distinguish between statistical fluctuations (upper-left plot) and genuine anomalies (all the other plots). 

\section{Conclusions}
In this work we propose a fast machine learning approach for data quality monitoring. 
The algorithm compares collected measurements with a reference dataset describing the standard detector readout performing a multidimensional likelihood-ratio hypothesis test. The significance of an eventual discrepancy in the data is quantified via a frequentist p-value. Some solutions for the location and characterization of the anomaly in the space of the input features are presented. Those could be relevant in order to design an automatic tool for the system control that allows to reset the apparatus to normal condition acting on the specific source of malfunctioning. The model is fast and can be used for quasi-online monitoring. The performances of the model can be tuned varying the size of the experimental data to be tested, the algorithm parameters and the computational resources. The experimental setup used to perform this study is simple but representative of the type of detectors installed in LHC experiments. A dedicated study tailoring more specific use cases will be performed in order to fit the requirements for the online deployment. 

\begin{ack}
M.L and M.R.  acknowledge the financial support of the European Research Council (grant SLING 819789).  G.G.  is supported by the European Research Council (ERC) under the European Union’s Horizon 2020 research and innovation program (grant agreement no 772369).  A.W. acknowledges support from the PRIN grant 2017FMJFMW.
\end{ack}

{
\small


}


\end{document}